\documentclass[12pt]{article}
\usepackage{amsmath}
\usepackage{graphicx, epsfig}
\usepackage{times}

\begin{document}
\begin{titlepage}
\title{The role of  double helicity-flip amplitudes in small-angle elastic
$pp$-scattering}
\author{S. M. Troshin, N. E. Tyurin\\[1ex]  Institute
for High Energy Physics,\\ Protvino, Moscow Region, 142281 Russia}
\date{}
\maketitle
\begin{abstract}
We asses a role of the double helicity-flip amplitudes in small-angle
elastic $pp$-scattering and
obtain a new unitary bound for the
double helicity-flip amplitude $F_2$ in
 elastic $pp$-scattering at small
values of $t$ on the basis of the $U$--matrix method of the $s$--channel
unitarization.\\[2ex] PACS number(s):  13.60.Hb, 13.88.+e
\end{abstract}
\vfill
\end{titlepage}

Discussion of a role and magnitude of helicity-flip amplitudes in small-angle
elastic scattering
has a long history  and is an important issue in
the studies of the spin properties of diffraction. Recently
an interest in  accounting the contributions
of single helicity-flip amplitudes becomes associated with
CNI polarimetry related problems \cite{but,kop,but1} as well.
Bound for the single helicity-flip amplitude
$F_5$ of  elastic $pp$-scattering
valid at finite energies has been derived in \cite{but}.
It corresponds to the asymptotic bound $cs\ln^3s$ for the function
$\hat{F}_5(s,0)\equiv[{mF_5(s,t)}/{\sqrt{-t}}]|_{t=0}$.
Asymptotic unitarity bound valid in high energy
limit  obtained in \cite{uf5} is stronger and it
 shows that $\hat{F}_5(s,0)$
cannot rise at
$s\rightarrow \infty$ faster  than
$cs\ln^2s$, i.e.
this bound is similar to the Froissart-Martin bound
for the helicity non-flip amplitudes.
However, not only non-flip and single helicity-flip amplitudes can give
contributions and affect  the estimates and bounds for the analyzing
power $A_N$. Double helicity-flip amplitudes can also contribute
into $A_N$ and  their behavior at high energies is also
 important  for the spin correlation
parameters and total cross-section differences in experiments with two
polarized beams available at RHIC nowadays.

The double helicity-flip amplitudes are usually
 neglected since they are supposed to be small
 in the whole region of momentum transfers. But this assumption  is based
merely on the technical simplification of the problem and is not
valid at large momentum transfers in elastic $pp$-scattering
where double-flip amplitudes can play an important role and fill up
multiple-dip structure in differential cross-section providing
correct description of the experimental data \cite{uf24}.
It is natural then to asses the role of double helicity-flip amplitudes
at small and moderate values of $t$ also.
In this note we use  unitarization method
 based on the $U$-matrix approach
 and obtain bounds for the amplitudes $F_2$ and $F_4$
which provide ground for the assumptions on their size and lead
to the high-energy bounds for the cross-section difference $\Delta\sigma_T(s)$.

The method is based on the
unitarity equation for helicity amplitudes of elastic
$pp$-scattering. It should be noted here that there
 is no  universal, generally accepted
method to implement unitarity in high energy scattering.
However, a choice of particular unitarization scheme is not completely a matter
of taste. Long time ago the arguments based on analytical properties of the scattering
amplitude  were put forward \cite{blan} in favor of the rational form of unitarization.
It was shown  that this form of unitarization reproduced
correct analytical properties of the scattering amplitude
in the complex energy plane much easier compared to the
exponential form, where simple  singularities of the eikonal function
would lead to the essential singularities in the amplitude.
 In potential scattering the eikonal (exponential)
and $U$--matrix (rational)
forms of unitarization correspond to two different approximations
of the scattering wave function, which satisfy
the Schr\"odinger equation to the same order \cite{blan}.
Rational form of unitarization
corresponds to an approximate wave function which changes both
the phase and amplitude of the wave. This form follows from dispersion
theory. It can be rewritten in the exponential
form but with completely different resultant phase function, and
relation of the two phase functions is given in \cite{blan}.
The rational form of unitarization in quantum field theory
 is based on the relativistic generalization \cite{umat}
 of the Heitler equation of radiation dumping \cite{heit}.
 In this approach an elastic scattering amplitude
 (we consider scattering of  spinless particles for the
 moment)  is a solution  of  the
following equation in the c.m.s.
\begin{equation}\label{xx}
 F({\bf p},{\bf q})  =
   U({\bf p},{\bf q})+
  i\frac{\pi}{8}\rho
(s)\int d\Omega_{{ \bf \hat k}}
U({\bf p},{\bf k})
F({\bf k},{\bf q}),
\end{equation}
where ${\bf p}={{\bf p}_1}=-{{\bf p}_2}$ and ${\bf q}={{\bf q}_1}=-{{\bf q}_2}$
are momenta of the initial and final particles. The kinematical
factor $\rho(s)\simeq 1$ at $s\gg 4m^2$ and
will be neglected in the following. The equation (\ref{xx}) has  simple solution
in the impact parameter representation\footnote{We factored out here an imaginary unity to
provide a more compact form for the helicity amplitudes in what following.}:
\[
(F,U)(s,t)=i\frac{s}{\pi^2}
\int_0^\infty bdb (f,u)(s,b)
J_{0}(b\sqrt{-t}),
\]
i.e.
\begin{equation}\label{rat}
f(s,b)=\frac{u(s,b)}{1+u(s,b)}.
\end{equation}
 Eq. (\ref{xx}) allows  one to fulfill the unitarity provided the
 inequality \begin{equation} \mbox{Re} u(s,b) \geq 0 \end{equation}
is satisfied\footnote{This is the only requirement needed to get an amplitude limited
by unity $|f(s,b)|\leq 1$ (as unitarity requires),
the function $u(s,b)$ itself should not obey such constraint.} .
The inelastic overlap function,
\[
\eta(s,b)\equiv\frac{1}{4\pi}\frac{d\sigma_{inel}}{db^2},
\]
i.e.  the sum of
all inelastic channel contributions into unitarity equation
\begin{equation}
\mbox{Re} f(s,b)=|f(s,b)|^2+\eta(s,b), \label{unt}
\end{equation}
has the following expression in terms of the function $u(s,b)$:
\begin{equation}
\eta(s,b)=\frac{\mbox{Re} u(s,b)}{|1+u(s,b)|^{2}}\label{uf}.
\end{equation}
The function $U(s,t)$ is the generalized reaction matrix,
which is considered to be an
input dynamical quantity similar to eikonal function. In potential scattering
this function  is related to the potential \cite{blan}, i.e.
\[
u(s,b)\sim \int_{-\infty}^\infty dz V(\sqrt{z^2+b^2}).
\]
 Construction of the  particular models for the relativistic case
  in the framework of the $U$--matrix
approach proceeds the common steps, i.e. the basic dynamics as
well as the notions on hadron structure being used  to obtain a
particular form for the $U$--matrix.
It is interesting to note that
the  form for the scattering amplitude analogous to Eq. (\ref{rat})
 was obtained  by Feynman in his
parton model for diffractive scattering
(which he has never published, cf. \cite{ravn}).

In what follows we will not use a model features and detailed structure
of $u(s,b)$,
 but consider  reasonable arguments
of a general nature, e.g.
for the function $u(s,b)$ we can adopt a simple form
\begin{equation}
u(s,b)=gs^\Delta e^{-\mu b},\label{usb}
\end{equation}
where the parameter $\Delta >0$ guarantees  the rise of the total cross-section.
  This is a rather general parameterization for $u(s,b)$ which
provides correct analytical properties in the complex $t$--plane, i.e.
it is
 consistent with
the representation  for the
function $u(s,b)$:
\begin{equation}\label{spec}
u(s,b)=\frac{\pi^2}{is}\int_{t_0}^\infty\omega(s,t)K_0(b\sqrt{t})dt.
\end{equation}
The Eq. (\ref{spec}) is a Fourier--Bessel transform of the spectral representation
for the $U$--matrix\footnote{In fact, it is valid separately for its
even and odd parts regarding cosine of the scattering angle.}:
\begin{equation}
U(s,t)=\int_{t_0}^\infty \frac{\omega(s,t')}{t'-t}dt',
\end{equation}
where the function $\omega(s,t)$ is the corresponding discontinuity of the
function $U(s,t)$ \cite{spect}.

Equation (\ref{xx}) for the helicity amplitudes
 of $pp$--scattering (i.e. for the two--fermion
scattering)
 has the following form
in the c.m.s. \cite{ech}:
\begin{eqnarray}
 F_{\lambda_3,\lambda_4,\lambda_1,\lambda_2}({\bf p},{\bf q}) & =
  & U_{\lambda_3,\lambda_4,\lambda_1,\lambda_2}({\bf p},{\bf q})+ \label{heq}\\
 & & i\frac{\pi}{8}
\sum_{\lambda ',\lambda ''}\int d\Omega_{{ \bf \hat k}}
U_{\lambda_3,\lambda_4,\lambda ',\lambda ''}({\bf p},{\bf k})
F_{\lambda ',\lambda '',\lambda_1,\lambda_2}({\bf k},{\bf q}),\nonumber
\end{eqnarray}
where
$\lambda 's$ are the initial and final proton's helicities. $F_i$ are the
helicity amplitudes in the standard notations, i.e.
\[
F_1\equiv F_{1/2,1/2,1/2,1/2}, \, F_2\equiv F_{-1/2,-1/2,1/2,1/2},
\, F_3\equiv F_{1/2,-1/2,1/2,-1/2}
\]
and
\[F_4\equiv F_{1/2,-1/2,-1/2,1/2},\, F_5\equiv F_{1/2,1/2,1/2,-1/2}.
\]

In the impact parameter representation for the helicity amplitudes $F_i$ and the
helicity functions $U_i$:
\[
(F,U)_{\lambda_3,\lambda_4,\lambda_1,\lambda_2}(s,t)=i\frac{s}{\pi^2}
(-1)^{N-\lambda}\int_0^\infty bdb (f,u)_{\lambda_3,\lambda_4,\lambda_1,\lambda_2}(s,b)
J_{|\lambda_1-\lambda_2-\lambda_3+\lambda_4|}(b\sqrt{-t}),
\]
where $N\equiv \min [(\lambda_1-\lambda_2),(\lambda_3-\lambda_4)]$,
$\lambda\equiv\lambda_1-\lambda_2$, we will have a system of the algebraic equations
\begin{equation}\label{heqb}
 f_{\lambda_3,\lambda_4,\lambda_1,\lambda_2}(s,b)  =
u_{\lambda_3,\lambda_4,\lambda_1,\lambda_2}(s,b) -  \sum_{\lambda ',\lambda ''}
u_{\lambda_3,\lambda_4,\lambda ',\lambda ''}(s,b)
f_{\lambda ',\lambda '',\lambda_1,\lambda_2}(s,b).
\end{equation}

Explicit solution of Eqs. (\ref{heqb}) then has the following
form:
\begin{eqnarray}
f_1 & = & \frac{(u_1 + u_1^2 - u_2^2)(1 + u_3 + u_4) - 2(1 + 2u_1 - 2u_2)u_5^2}
{(1 + u_1 - u_2)[(1 + u_1 + u_2)(1 + u_3 + u_4)- 4u_5^2]},\nonumber\\
f_2 & = & \frac{u_2(1 + u_3 + u_4) - 2u_5^2}
{(1 + u_1 - u_2)[(1 + u_1 + u_2)(1 + u_3 + u_4)- 4u_5^2]},\nonumber\\
f_3 & = & \frac{(u_3 + u_3^2 - u_4^2)(1 + u_1 + u_2) - 2(1 + 2u_3 - 2u_4)u_5^2}
{(1 + u_3 - u_4)[(1 + u_1 + u_2)(1 + u_3 + u_4)- 4u_5^2]},\nonumber\\
f_4 & = & \frac{u_4(1 + u_1 + u_2) - 2u_5^2}
{(1 + u_3 - u_4)[(1 + u_1 + u_2)(1 + u_3 + u_4)- 4u_5^2]},\nonumber\\
f_5 & = & \frac{u_5}
{(1 + u_1 + u_2)(1 + u_3 + u_4)- 4u_5^2},\label{fi}
\end{eqnarray}
where for simplicity we omitted in the functions $f_i(s,b)$
and $u_i(s,b)$ their arguments. Unitarity requires that
$\mbox{Re}u_{1,3}(s,b)\geq 0$, but the absolute values of the
functions $u_i(s,b)$ should not be limited by unity.
For the functions $u_{2,4}(s,b)$ we adhere to a simple general
 form similar to the above Eq. (\ref{usb}) (using arguments
 based on the analytical properties in the complex $t$--plane):
\begin{equation}
u_{2}\sim u_4 \sim s^{\Delta} e^{-\mu b}.\label{usbn}
\end{equation}
To get an upper bound for the amplitudes $F_{2,4}(s,t)$
we consider the case when $u_{2,4}(s,b)$ are dominating ones.
Then we have for the amplitudes $F_{2,4}(s,t)$ the following
representation
\begin{equation}
F_{2}(s,t)=\frac{is}{\pi^2}\int_0^\infty bdb \frac{u_{2}(s,b)}
{1-u_{2}^2(s,b)}J_0(b\sqrt{-t})\label{f2s}
\end{equation}
and
\begin{equation}
F_{4}(s,t)=\frac{is}{\pi^2}\int_0^\infty bdb \frac{u_{4}(s,b)}
{1-u_{4}^2(s,b)}J_2(b\sqrt{-t})\label{f4s}
\end{equation}
Using for $u_{2,4}(s,b)$ the functional dependence in the form of Eq.
(\ref{usb}) it can be  shown that  the amplitude $F_2(s,t=0)$
cannot rise faster than $s\ln s$ at $s\to\infty$ and the function
\[
\hat F_4(s,t=0)\equiv [\frac{m^2}{-t}F_4(s,t)]|_{t=0}
\]
cannot rise faster than $s\ln^3 s$ at $s\to\infty$.

Thus, we can state that the explicit account of unitarity in the
form of $U$ - matrix approach
leads to the following upper bound for the cross-section difference
\[
\Delta\sigma_T \leq c\ln s,
\]
where
\[
\Delta\sigma_T\equiv\sigma_{tot}(\uparrow\downarrow )-\sigma_{tot}(\uparrow\uparrow )
\sim-\frac{1}{s}\mbox{Im}F_2(s,t=0).
\]
It should be noted that the asymptotic behaviour of the amplitudes
$F_1$ and $F_3$ are determined by the functions $u_2$ and $u_4$, respectively, in the
situation when these functions dominate;  the
Froissart--Martin asymptotical bound  for these amplitudes remains under these
circumstances,  i.e. they are
limited by $cs\ln^2s$ at $t=0$.

Another related important  consequence is the conclusion on the possibility to neglect
helicity-flip amplitudes $F_2$, $F_4$ and $F_5$ under calculations of differential
cross-section
\[
\frac{d\sigma}{dt}=\frac{2\pi^5}{s^2}(|F_1(s,t)|^2+|F_2(s,t)|^2+|F_3(s,t)|^2+
|F_4(s,t)|^2+4|F_5(s,t)|^2)
\]
and double helicity-flip amplitudes $F_2$ and $F_4$ under calculation
of analyzing power $A_N$
\[
A_N(s,t)\frac{d\sigma}{dt}=\frac{2\pi^5}{s^2}\mbox{Im}[(F_1(s,t)+F_2(s,t)+F_3(s,t)
-F_4(s,t))^*F_5(s,t)]
\]
in the region of small values of $t$ in high energy limit. This conclusion is based
on the above bounds for the helicity amplitudes and their small $t$ dependence due to
angular momentum conservation, i.e.
at $-t\to 0$: $F_i \sim \mbox{const}$, $(i=1,2,3)$, $F_5\sim\sqrt{-t}$ and
$F_4\sim -t$. However, the dominance of the helicity-non-flip amplitudes ceases
to be valid at fixed values of momentum transfers, where , e.g. amplitude $F_4$ can
become a dominant one, since its energy growth is limited by the function $s\ln^3 s$, while
other helicity amplitudes cannot increase faster than $s\ln^2 s$.

One should recall that unitarity for the helicity amplitudes  leads to    a peripheral
dependence of the amplitudes $f_i(s,b)$ $(i=2,4,5)$
on the impact parameter $b$ at high energy, i.e.
\[
|f_i(s,b=0)|\rightarrow 0
\]
at $s\rightarrow\infty$. This is a consequence of the explicit
 unitarity
representation for the helicity amplitudes through the  $U$-matrix and it is this
fact allows
one to get better bounds for the helicity-flip amplitudes.

 Thus,  as it was shown in this note and in \cite{uf5}, we have
the following asymptotic results:
\begin{itemize}
\item
 the ratio
$r_5(s,0)\equiv2\hat{F}_5(s,0)/[F_1(s,0)+F_3(s,0)]$
cannot increase with energy,
\item
the amplitude $F_2(s,t=0)$
cannot increase faster than $s\ln s$,
\item
 the function
$\hat F_4(s,t=0)$
should not rise faster than $s\ln^3 s$ at high energies.
\end{itemize}
Nowadays
RHIC spin program includes experiments with two polarized proton beams at the
highest available energies and the above bounds could be useful
and provide grounds for the estimations of the spin observables
in the forward region in these experiments.
 The above bounds provide justification of
the smallness of the double helicity-flip amplitudes in the low-$t$ region, but
simultaneously they  imply an importance of the double helicity-flip
amplitudes at the moderate values of momentum transfers. This result is in accordance
with early analysis of experimental data performed in \cite{uf24}.
Magnitude of the helicity amplitude $F_2$ at $t=0$ can be measured directly
at RHIC through the measurements of $\Delta\sigma_T$ \cite{pp2pp}
 and it is definitely  an
important study of the spin properties of diffraction.
The experimental data for
$\Delta\sigma _T(s)$
could also be a useful source of information
on the low-$x$ behaviour of the spin structure function $h_1(x)$.

\end{document}